\documentclass[journal]{IEEEtran}
\makeatletter

\usepackage{blindtext}
\usepackage[T1]{fontenc}
\usepackage{textcomp}
\usepackage{amsmath}
\usepackage{amssymb}
\usepackage{bm}
\usepackage{mdwmath}
\usepackage{cases}
\usepackage{graphicx}
\graphicspath{{Figure/PDF/}{Biography/PDF/}}
\usepackage[dvipsnames]{xcolor}
\definecolor{myblue}{rgb}{0,0.4980,1} 
\definecolor{myred}{rgb}{0.8706,0.1608,0.0627} 
\usepackage[caption=false,font=footnotesize,subrefformat=parens]{subfig}
\usepackage[nosort,noadjust]{cite}
\usepackage{url}
\usepackage{tabularray}
\UseTblrLibrary{varwidth}
\UseTblrLibrary{diagbox}
\ExplSyntaxOn
\tl_const:Nn \c__tblr_valign_M_tl {M}
\tl_new:N    \g__tblr_cell_valign_sub_tl

\keys_define:nn { tblr-column }
  {
    M .meta:n = { valign = M },
  }

\cs_gset_protected:Npn \__tblr_get_cell_alignments:nn #1 #2
  {
    \group_begin:
    \tl_gset:Nx \g__tblr_cell_halign_tl
      { \__tblr_data_item:neen { cell } {#1} {#2} { halign } }
    \tl_set:Nx \l__tblr_v_tl
      { \__tblr_data_item:neen { cell } {#1} {#2} { valign } }
    \tl_case:NnF \l__tblr_v_tl
      {
        \c__tblr_valign_t_tl
          {
            \tl_gset:Nn \g__tblr_cell_valign_tl {m}
            \tl_gset:Nn \g__tblr_cell_middle_tl {t}
            \tl_gclear:N \g__tblr_cell_valign_sub_tl
          }
        \c__tblr_valign_m_tl
          {
            \tl_gset:Nn \g__tblr_cell_valign_tl {m}
            \tl_gset:Nn \g__tblr_cell_middle_tl {m}
            \tl_gclear:N \g__tblr_cell_valign_sub_tl
          }
        \c__tblr_valign_M_tl
          {
            \tl_gset:Nn \g__tblr_cell_valign_tl {m}
            \tl_gset:Nn \g__tblr_cell_valign_sub_tl {M}
            \tl_gset:Nn \g__tblr_cell_middle_tl {t}
          }
        \c__tblr_valign_b_tl
          {
            \tl_gset:Nn \g__tblr_cell_valign_tl {m}
            \tl_gset:Nn \g__tblr_cell_middle_tl {b}
            \tl_gclear:N \g__tblr_cell_valign_sub_tl
          }
      }
      {
        \tl_gset_eq:NN \g__tblr_cell_valign_tl \l__tblr_v_tl
        \tl_gclear:N \g__tblr_cell_middle_tl
      }
    \group_end:
  }

\cs_gset_protected:Npn \__tblr_build_cell_content:NN #1 #2
  {
    \hbox_set_to_wd:Nnn \l__tblr_a_box { \l__tblr_cell_wd_dim }
      {
        \tl_if_eq:NnTF \g__tblr_cell_halign_tl {j}
          { \__tblr_get_cell_text:nn {#1} {#2} \hfil }
          {
            \tl_if_eq:NnF \g__tblr_cell_halign_tl {l} { \hfil }
            \__tblr_get_cell_text:nn {#1} {#2}
            \tl_if_eq:NnF \g__tblr_cell_halign_tl {r} { \hfil }
          }
      }
    \vbox_set_to_ht:Nnn \l__tblr_b_box { \l__tblr_cell_ht_dim }
      {
        \tl_case:Nn \g__tblr_cell_valign_tl
          {
            \c__tblr_valign_m_tl
              {
                \vfil
                \bool_lazy_and:nnT
                  { \int_compare_p:nNn { \l__tblr_cell_rowspan_tl } < {2} }
                  { !\tl_if_eq_p:NN \c__tblr_valign_M_tl \g__tblr_cell_valign_sub_tl }
                  {
                    \box_set_ht:Nn \l__tblr_a_box
                      { \__tblr_data_item:nen { row } {#1} { @row-upper } }
                    \box_set_dp:Nn \l__tblr_a_box
                      { \__tblr_data_item:nen { row } {#1} { @row-lower } }
                  }
                \box_use:N \l__tblr_a_box
                \vfil
              }
            \c__tblr_valign_h_tl
              {
                \box_set_ht:Nn \l__tblr_a_box
                  { \__tblr_data_item:nen { row } {#1} { @row-head } }
                \box_use:N \l__tblr_a_box
                \vfil
              }
            \c__tblr_valign_f_tl
              {
                \vfil
                \int_compare:nNnTF { \l__tblr_cell_rowspan_tl } < {2}
                  {
                    \box_set_dp:Nn \l__tblr_a_box
                      { \__tblr_data_item:nen { row } {#1} { @row-foot } }
                  }
                  {
                    \box_set_dp:Nn \l__tblr_a_box
                      {
                        \__tblr_data_item:nen
                          { row }
                          { \int_eval:n { #1 + \l__tblr_cell_rowspan_tl - 1 } }
                          { @row-foot }
                      }
                  }
                \box_use:N \l__tblr_a_box
              }
          }
        \hrule height ~ 0pt 
      }
    \vbox_set_to_ht:Nnn \l__tblr_c_box
      { \l__tblr_row_ht_dim - \l__tblr_row_abovesep_dim }
      {
        \box_use:N \l__tblr_b_box
        \vss
      }
    \skip_horizontal:n { \l__tblr_x_tl }
    \box_use:N \l__tblr_c_box
    \skip_horizontal:n { \l__tblr_y_tl - \l__tblr_cell_wd_dim + \l__tblr_w_tl }
  }
\ExplSyntaxOff

\usepackage{hyperref}
\newcommand{\colorhypersetup}{\@ifpackageloaded{hyperref}{\hypersetup{%
	bookmarksopen=true,%
	bookmarksnumbered=true,%
	pdfpagemode={UseOutlines},
	pdfstartview={FitH},%
	colorlinks=true,%
	linkcolor={myred},%
	citecolor={orange}
}}{\empty}}
\newcommand{\blackhypersetup}{\@ifpackageloaded{hyperref}{\hypersetup{%
	bookmarksopen=true,%
	bookmarksnumbered=true,%
	pdfpagemode={UseOutlines},
	pdfstartview={FitH},%
	colorlinks=true,%
	allcolors={black}
}}{\empty}}
\blackhypersetup

\usepackage{mfirstuc-english}
\MFUhyphentrue
\usepackage[version3,description,IEEEtran]{eacro}
\acsetup{%
    pages/display=all,%
    pages/seq/use=false,%
    pages/name=true,%
    pages/fill={\quad},%
    make-links=false,%
    case-sensitive=false,
}
\DeclareAcronym{vr}{
    short = VR,
    long = virtual reality}
\DeclareAcronym{3d}{
    short = 3D,
    long = three-dimension}
\DeclareAcronym{2d}{
    short = 2D,
    long = two-dimension}
\DeclareAcronym{6g}{
    short = 6G,
    long = sixth-generation}
\DeclareAcronym{embb}{
    short = eMBB,
    long = enhanced mobile broadband}
\DeclareAcronym{ai}{
    short = AI,
    long = artificial intelligence}
\DeclareAcronym{dt}{
    short = DT,
    long = digital twin}
\DeclareAcronym{VDTNet}{
    short = DTN4VS,
    long = \acs*{dt}-driven network architecture for video streaming}
\DeclareAcronym{hd}{
    short = HD,
    long = high-definition}
\DeclareAcronym{uhd}{
    short = UHD,
    long = ultra-high-definition}
\DeclareAcronym{nc}{
    short = NC,
    long = network controller}
\DeclareAcronym{roi}{
    short = RoI,
    long = region of interest}
\DeclareAcronym{qos}{
    short = QoS,
    long = quality of service}
\DeclareAcronym{qoe}{
    short = QoE,
    long = quality of experience}
\DeclareAcronym{6dof}{
    short = 6-DoF,
    long = six degrees of freedom}
\DeclareAcronym{gnn}{
    short = GNNs,
    long = graph neural networks}
\DeclareAcronym{drl}{
    short = DRL,
    long = deep reinforcement learning}
\DeclareAcronym{gan}{
    short = GAN,
    long = generative adversarial networks}
\DeclareAcronym{kpi}{
    short = KPI,
    long = key performance indicator}
\DeclareAcronym{cdn}{
    short = CDN,
    long = content delivery network}
\DeclareAcronym{udt}{
    short = UDT,
    long = user \acs*{dt}}
\DeclareAcronym{idt}{
    short = IDT,
    long = infrastructure \acs*{dt}}
\DeclareAcronym{sdt}{
    short = SDT,
    long = slice \acs*{dt}}
\DeclareAcronym{bs}{
    short = BS,
    long = base station}
\DeclareAcronym{voi}{
    short = VoI,
    long = value of information}
\DeclareAcronym{vgg}{
    short = VGG,
    long = visual geometry group}
\DeclareAcronym{vit}{
    short = ViT,
    long = vision transformers}
\DeclareAcronym{iot}{
    short = IoT,
    long = Internet of Things}
\DeclareAcronym{fov}{
    short = FoV,
    long = field of view}
\DeclareAcronym{es}{
    short = ES,
    long = edge server}
\DeclareAcronym{uw}{
    short = UW,
    long = University of Waterloo}

\usepackage{algpseudocode}
\newcounter{MYalgorithmic}

{%
	\vspace*{3pt}
	\hrule height 1pt}
\newcounter{MYitem}[MYalgorithmic]

\newcommand{\MYlabel}[1]{\def\@currentlabel{\theALG@line}\label{#1}}

\newcommand{\upperroman}[1]{\uppercase\expandafter{\romannumeral#1}}

\newcommand{\myvec}[1]{\bm{\mathrm{#1}}}

\newcommand{\myunit}[1]{%
	\ifmmode
		\mathrm{#1}
	\else
		$ \mathrm{#1} $
	\fi}
\newcommand{\murm}{%
	\ifmmode
		\text{\textmu}
	\else
		\textmu
	\fi}

\newcommand{\MYnewpage}{%
	\ifCLASSOPTIONonecolumn
		\ifCLASSOPTIONjournal
			\typeout{The onecolumn journal mode.}
			\newpage
		\fi
	\fi}

\newlength{\mysinglefigwidth}
\newlength{\mymultifigwidth}
\ifCLASSOPTIONonecolumn
	\AtBeginDocument{\setlength{\mysinglefigwidth}{0.5\linewidth}}
\else
	\AtBeginDocument{\setlength{\mysinglefigwidth}{\linewidth}}
\fi
\ifCLASSOPTIONonecolumn
	\AtBeginDocument{\setlength{\mymultifigwidth}{0.5\linewidth}}
\else
	\AtBeginDocument{\setlength{\mymultifigwidth}{\linewidth}}
\fi

\makeatother
\begin{document}
\title{Digital Twin-Driven Network Architecture for Video Streaming}

\author{Xinyu~Huang,~\IEEEmembership{Student~Member,~IEEE}, Haojun~Yang,~\IEEEmembership{Member,~IEEE}, Shisheng~Hu,~\IEEEmembership{Student~Member,~IEEE}, and~Xuemin~(Sherman)~Shen,~\IEEEmembership{Fellow,~IEEE}
    \thanks{Xinyu Huang, Haojun Yang, Shisheng Hu, and Xuemin (Sherman) Shen are with the Department of Electrical and Computer Engineering, University of Waterloo, Waterloo, ON, N2L 3G1, Canada (E-mail: \{x357huan, haojun.yang, s97hu, sshen\}@uwaterloo.ca). {Corresponding author:} Haojun Yang.}
}

\ifCLASSOPTIONonecolumn
	\typeout{The onecolumn mode.}
\else
	\typeout{The twocolumn mode.}
	\markboth{{To be Published in IEEE Network}}{Author \MakeLowercase{\textit{et al.}}: Title}
\fi

\maketitle

\ifCLASSOPTIONonecolumn
	\typeout{The onecolumn mode.}
	\vspace*{-50pt}
\else
	\typeout{The twocolumn mode.}
\fi
\begin{abstract}

Digital twin (DT) is revolutionizing the emerging video streaming services through tailored network management. By integrating diverse advanced communication technologies, DTs are promised to construct a holistic virtualized network for better network management performance. To this end, we develop a DT-driven network architecture for video streaming (DTN4VS) to enable network virtualization and tailored network management. With the architecture, various types of DTs can characterize physical entities' status, separate the network management functions from the network controller, and empower the functions with emulated data and tailored strategies. To further enhance network management performance, three potential approaches are proposed, i.e., domain data exploitation, performance evaluation, and adaptive DT model update. We present a case study pertaining to DT-assisted network slicing for short video streaming, followed by some open research issues for DTN4VS.
\end{abstract}

\ifCLASSOPTIONonecolumn
	\typeout{The onecolumn mode.}
	\vspace*{-10pt}
\else
	\typeout{The twocolumn mode.}
\fi


\IEEEpeerreviewmaketitle

\MYnewpage

\section{Introduction}

\acresetall

Video streaming services have evolved dramatically, transitioning from simple on-demand platforms to sophisticated and real-time interactive systems. A statistics report shows that the global video streaming market size was valued at \$$89$ billion in 2022, and is expected to grow at a compound annual growth rate of $21.5$\% until 2030~\cite{report}. Emerging video streaming services, including intelligent short video streaming, extreme immersive \ac{vr}, and holographic video streaming, demand tailored network management to satisfy users' personalized requirements~\cite{9599635}. For instance, by adding multiple video branches and view angles, intelligent short video streaming emphasizes the interaction with users, which is triggered by the users' swipe and rotation behaviors. To satisfy smooth video playback while reducing bandwidth consumption, communication networks should accurately mine the preferences and behavior characteristics of users for better intelligent video caching. Furthermore, extreme immersive \ac{vr} and holographic video streaming aim at providing high-fidelity \ac{3d} object display and immersive experience, which demand efficient coordination between sensing, video tile transmission, video rendering, and specialized video codecs. To satisfy these evolving requirements, efficient network management through advanced communication technologies becomes an imperative endeavor.

Advanced communication technologies, such as \ac{embb}-Plus, native \ac{ai}, sensing, network slicing, and \ac{dt}, are expected to satisfy the above requirements~\cite{dang2020should,9267778}. For instance, \ac{embb}-Plus can provide gigabit-level data rates and seamless connections, while native \ac{ai} enables intelligent data processing and decision-making. Moreover, sensing-related techniques facilitate real-time \ac{3d} object modeling, and network slicing is used to isolate network resources for prescribed service requirements. To seamlessly integrate these technologies for video streaming services, a holistic network management architecture is essential. As a promising approach, \ac{dt} can realize the holistic network virtualization for video streaming services by exploiting its real-time monitoring, analytics, and emulation capabilities. Specifically, \acp{dt} can characterize users' real-time status, \ac{qos}, and \ac{qoe} through native \ac{ai} and sensing, and provide an emulation environment for network management by implementing tailored network slicing and resource allocation policies on \ac{embb}-Plus. By leveraging the capabilities of \acp{dt}, an efficient holistic network management architecture for video streaming services can be realized.

However, developing an efficient \ac{dt}-driven network architecture for video streaming services faces many challenges, such as
\begin{itemize}
\item \textit{Lack of Efficient Data Abstraction Mechanism:} An efficient data abstraction mechanism should be developed to facilitate the real-time and intricate interplay among \acp{dt}, slice domain, and physical domain, which includes the determination of data types, granularities, and features.

\item \textit{Lack of Comprehensive Performance Evaluation Framework:} Since the \ac{dt} performance consists of two-fold aspects, i.e., the accuracy and cost of itself, and its impact on network performance, it is crucial to develop a new and comprehensive performance evaluation framework to evaluate the \ac{dt} performance.

\item \textit{Lack of Adaptive \ac{dt} Model Update:} Due to the distinct spatiotemporal dynamics that exist in network conditions and user behaviors, as well as diversified service requirements, it is essential to fine-tune tailored \ac{dt} models to adapt to the dynamics and diversity.
\end{itemize}

In this article, we propose a \ac{VDTNet} to enable network virtualization and tailored network management. Three kinds of DTs, i.e., \ac{udt}, \ac{idt}, and \ac{sdt}, are built to characterize the network from the user-level, operation-level, and service-level perspectives, respectively. \acp{dt} can provide distilled user information, emulated environment, and tailored network management strategies to realize efficient network management. To tackle the mentioned challenges, we first propose an efficient data collection, fusion, and abstraction mechanism. Secondly, we propose a comprehensive performance evaluation framework, which integrates \ac{dt} data freshness, \ac{qos}/\ac{qoe} gain, and \ac{dt} operation cost. Thirdly, we propose an adaptive \ac{dt} model update method that integrates distributed and transfer learning algorithms to realize computing load balance and computing overhead reduction. A case study pertaining to \ac{dt}-assisted network slicing for short video streaming is presented, followed by a discussion on potential research issues.

The remainder of this article is organized as follows. Firstly, emerging video streaming services and the corresponding communication techniques are discussed, followed by the proposed \ac{VDTNet}. Then, we discuss the challenges for \ac{VDTNet} and some potential solutions. Next, a case study about DT-assisted network slicing for short video streaming is presented. Finally, the open research issues are identified, followed by the conclusion.

\section{DT-Driven Network Architecture for Video Streaming}

In this section, we first introduce emerging video streaming services and advanced communication technologies, and then the \ac{VDTNet} architecture is proposed.

\subsection{Emerging Video Streaming}
Various innovative video streaming is emerging, including intelligent short video streaming, extreme immersive \ac{vr} streaming, and holographic video streaming, which poses enhanced requirements on communication networks, as shown in Table~\ref{tab1}.

\subsubsection{Intelligent Short Video Streaming}

It has two main characteristics, i.e., rotation-based swipe and multi-branch. The front indicates that the short video streaming will be extended from the current \ac{2d} format to \ac{3d} format, thus users can watch different angles through rotation-based swipe behaviors. Video tiles are selectively transmitted to user terminals based on the analysis of users’ swipe behaviors to reduce network traffic load. The latter means that more video branches will be added to the main video storyline to boost users' interaction. Part of critical video branches are preferentially cached in users' buffers to save bandwidth consumption.

\subsubsection{Extreme Immersive \ac{vr} Streaming}

It provides immersive and interactive experience by transmitting real-time \ac{3d} videos and audio to specialized headsets or mobile devices. Since future \ac{vr} streaming is expected to provide a panoramic view and ultra-high-definition resolution, field of view (FoV) transmission is an effective method to reduce the transmission burden. Furthermore, the latency requirements of extreme immersive \ac{vr} are very stringent, reaching just tens of milliseconds. It is essential to develop advanced sensing technologies to quickly capture users' dynamic behaviors, and optimize the computing process for video rendering.

\subsubsection{Holographic Video Streaming}

It refers to the \ac{3d} holographic content transmission, which usually requires the sensing-assisted communication technology to perceive users' behaviors and conduct \ac{3d} object modeling. Due to the ultra-low delay and strong interaction, high-performance computing nodes are needed to quickly compress and render the holographic video. For instance, to swiftly respond to users' movements in the \ac{6dof},  the end-to-end delay needs to be lower than 5~$ms$~\cite{9915358}.

\begin{table*}[!t]
\renewcommand{\IEEEiedlistdecl}{\setlength{\IEEElabelindent}{0pt}}
\centering
\caption{Emerging Video Streaming Services}
\label{tab1}
\begin{tblr}{
    width = 0.95\linewidth,
    colspec = {X[0.7,c,M]X[0.8,l,m]X[0.8,c,m]X[1.5,c,m]X[2,l,m]X[0.7,c,m]},
    hlines,
    hline{2} = {1}{-}{},
    hline{2} = {2}{-}{},
    vline{2-6},
    row{1} = {c,font=\bfseries},
    column{1} = {font=\bfseries},
    columns = {rightsep=3pt},
    cell{3}{4} = {l},
    cell{2}{5} = {c},
    measure=vbox,
}
Video Type & Characteristics & Video Codec & Bandwidth Requirements & Latency Requirements & Component \\ 
Intelligent Short Video & \begin{itemize}
    \item Swipe
    \item Multi-Branch
\end{itemize} & H. 264, MPEG & 4K: $45~\myunit{Mbps}$ & Several Seconds & Segments, \ac{3d} Tiles \\ 
Extreme Immersive \ac{vr} Video & \begin{itemize}
    \item Interactive
    \item Viewpoint
    \item Rendering
\end{itemize} & X\ac{3d}, MPEG-I~\cite{8258607} & \begin{itemize}
    \item 8K: $80\sim100~\myunit{Mbps}$
    \item 30K: $800\sim1000~\myunit{Mbps}$
\end{itemize} & \begin{itemize}
    \item Strong Interaction Mode: $5\sim10~\myunit{ms}$
    \item Weak Interaction Mode: $10\sim20~\myunit{ms}$
\end{itemize} & \ac{2d}/\ac{3d} Tiles \\ 
Holographic Video & \begin{itemize}
    \item Interactive
    \item \ac{3d} Modeling
    \item Rendering
\end{itemize} & HEVC, AV1, VP9 & 4K: $100~\myunit{Mbps}$ & \begin{itemize}
    \item \ac{6dof} Movement: $5~\myunit{ms}$
    \item 3-DoF Movement: $20~\myunit{ms}$
\end{itemize} & \ac{3d} Tiles 
\end{tblr}
\end{table*}

\subsection{Advanced Communication Techniques for Video Streaming}
In the rapidly evolving landscape of emerging video streaming, advanced communication techniques, such as eMBB-Plus, native AI, sensing, network slicing, and DT, can offer groundbreaking solutions to satisfy enhanced requirements, including personalized video buffering, ultra-low-latency interaction, and smooth \ac{3d} video transmission.
\subsubsection{\acs*{embb}-Plus}
As a cornerstone technology, eMBB-Plus is designed to provide higher bandwidth capacity, wider access coverage, and smarter caching and computing, which can effectively improve network throughput and reduce rebuffering. 
\begin{itemize}
    \item 
For intelligent short video streaming, eMBB-Plus can provide efficient distributed video caching and collaborative transcoding strategies to ensure seamless delivery of \ac{uhd} video segments.
\item 
For extreme immersive \ac{vr} streaming, eMBB-Plus can provide the increased frequency band and exploit more advanced modulation and coding techniques to ensure the smooth transmission for high-bitrate \ac{3d} videos.
\item
In holographic video streaming, since \ac{3d} modeling occupies plenty of computing time, eMBB-Plus will provide more advanced data offloading and collaborative computing mechanisms to help reduce computing delay.
\end{itemize}

\subsubsection{Native AI}
As a built-in component in next-generation communication networks, native AI is promised to provide more intelligent data processing and resource management strategies~\cite{9749222}.
\begin{itemize}
    \item 
    For intelligent short video streaming, \ac{gnn} can model the complex relationship between users and contents to enable more personalized and context-aware video recommendations and buffering.
    \item 
     For extreme immersive \ac{vr} streaming, \ac{drl} algorithms can accurately allocate network resources and distribute video tiles to adapt to users' dynamic behaviors for smooth video playback.
     \item 
     In holographic video streaming, convolutional neural networks (CNNs) can be employed for efficient data compression and semantic extraction, significantly reducing bandwidth consumption while maintaining a high fidelity.
\end{itemize}

\subsubsection{Sensing}

It can capture users' real-time macro and micro behaviors to help tailor network management for individual users.
\begin{itemize}
\item For intelligent short video streaming, advanced facial recognition sensors can capture users' micro-expressions, which are further analyzed by machine learning algorithms for intelligent video recommendations and buffer control. 

\item In extreme immersive \ac{vr} streaming, communication and sensing signals can be multiplexed in the time, frequency, and spatial domains to improve spectrum utilization. For instance, IEEE 802.11 working group proposed the Wi-Fi sensing technology to exploit the features of the physical layer and medium access control, which can measure users' motion in real time~\cite{9941042}.

\item For holographic video streaming, efficient radio spectrum allocation and advanced beamforming technologies can effectively improve both data transmission reliability and holographic video resolution~\cite{9363029}.
\end{itemize}

\subsubsection{Network Slicing}
Empowered by technologies such as software-defined networking (SDN) and network function virtualization (NFV), network slicing can provide isolated resources for emerging video streaming services to satisfy the differentiated requirements~\cite{9749222}.
\begin{itemize}
    \item 
For intelligent short video streaming, network resources can be sliced to support real-time analysis of video content and user behaviors, thereby enabling intelligent video recommendation and adaptive video delivery.
\item 
In the realm of extreme immersive \ac{vr}, specialized slices can be constructed for high-performance sensors and computing nodes to satisfy ultra-low-latency requirements.
\item 
For holographic video streaming, dedicated slices can guarantee high bandwidth and computing requirements, facilitating real-time and high-fidelity \ac{3d} interactions.

\end{itemize}

\subsubsection{Digital Twin}

It is defined as a digital representation of a physical object or a process and real-time synchronization between the physical object or process~\cite{9651548}. DT is usually classified into three types, i.e., \ac{udt}, \ac{idt}, and \ac{sdt}, deployed at the core network and network edge nodes.
\begin{itemize}
    \item 
\ac{udt} corresponds to an end user that reflects its fine-grained information, such as network conditions, playback status, and interaction behaviors, etc. \ac{udt}s can emulate user status to help the \ac{nc} make tailored resource management strategies.
\item 
\ac{idt} is a digital mirror of network infrastructure, such as a \ac{bs} or an edge server, which reflects its operation status, traffic load, resource utilization, etc. \acp{idt} can separate the resource operation function from the \ac{nc} and empower the function with emulated data and tailored strategies.
\item 
\ac{sdt} is constructed by aggregating \ac{udt}s and \ac{idt}s to obtain coarse-grained distilled information, such as spatiotemporal service demand distribution, resource utilization, etc. By separating the resource planning function from the \ac{nc}, \acp{sdt} can strengthen the function's capability with emulated data and tailored strategies.
\end{itemize}
Based on fine- and coarse-grained information and tailored network management strategies via \ac{dt}, customized network resources are allocated to users to enhance user experience.

\subsection{\acs*{VDTNet}}

\begin{figure*}[!t]
	\centering
	\includegraphics[width=\textwidth]{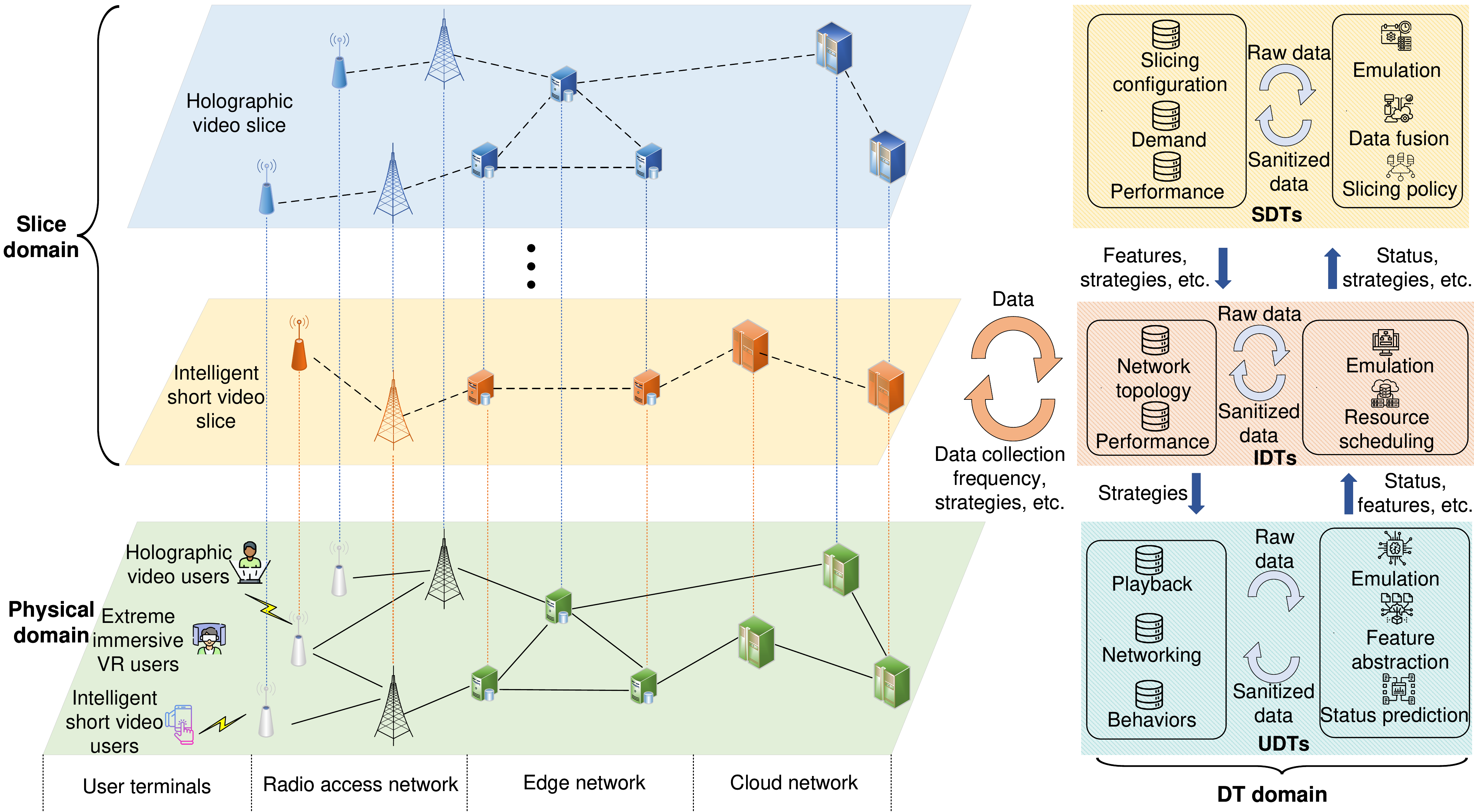}
	\caption{\ac{VDTNet} framework.}
	\label{frame}
\end{figure*}

To seamlessly integrate these emerging technologies, as shown in Fig.~\ref{frame}, we develop a \ac{VDTNet} framework to enhance video streaming service performance. The physical domain includes real-world video streaming infrastructures, while the slice domain leverages network slicing for prescribed \ac{qos}. The DT domain provides real-time digital replicas for data analytics, emulation, and network management decision-making, where an orchestration among \acp{udt}, \acp{idt}, and \acp{sdt} is intelligently coordinated to facilitate efficient network management.

\subsubsection{Physical Domain}

In the physical domain, the proposed \ac{VDTNet} meticulously integrates user terminals, radio access networks (RANs), edge networks, and cloud networks to build a holistic video streaming ecosystem. User terminals are not merely endpoints for video delivery but are also equipped with advanced codec technologies and adaptive bitrate (ABR) algorithms for efficient data exchange with RANs. The RAN layer consists of small \acp{bs} (S\acp{bs}) and macro \acp{bs} (M\acp{bs}) that employ advanced communication technologies, such as eMBB-Plus and sensing, to support high-throughput and high-fidelity video transmission. Edge networks provide localized computing and caching capabilities for latency-sensitive video streaming, while cloud networks handle large-scale video processing and storage. These elements collectively form the backbone of our framework, and through real-time resource scheduling and optimization in the DT domain, we can achieve holistic network management.

\subsubsection{Slice Domain}

Network slicing emerges as a pivotal technology for guaranteeing \ac{qos} requirements. A single physical network can be partitioned into multiple isolated slices to satisfy differentiated service requirements. For instance, one slice could be optimized for low-latency and interactive \ac{vr} video streaming, while another might target high-throughput holographic video streaming. Each slice consists of a unique set of network resources, policies, and protocols, which can enable tailored control over network resources. Through intelligent orchestration in the DT domain, these slices can be dynamically adjusted to meet varying resource demands, thereby achieving a harmonious balance between resource utilization and service quality. 

\subsubsection{DT Domain}

In the DT domain, each kind of DTs consists of a finite database and a model pool. For instance, the database of \acp{udt} includes users' playback-related, network-related, and behavior-related data. These data can reflect users' actual watching process, and be analyzed to fine-tune native AI models, such as long short-term memory (LSTM), recurrent neural network (RNN), and CNN, etc., to emulate and predict user status, and abstract distilled features. As a crucial hub connecting \acp{udt} and \acp{sdt}, \acp{idt} can further aggregate \acp{udt}' data to obtain some global information that can be provided to \acp{sdt} for slice adjustment. Furthermore, \acp{idt} is responsible for interacting with the physical domain and slice domain, such as adaptive data collection frequency and resource management strategies, etc.

\subsection{A Processing Procedure Example for Short Video}

\begin{figure}[!t]
	\centering
	\includegraphics[width=\linewidth]{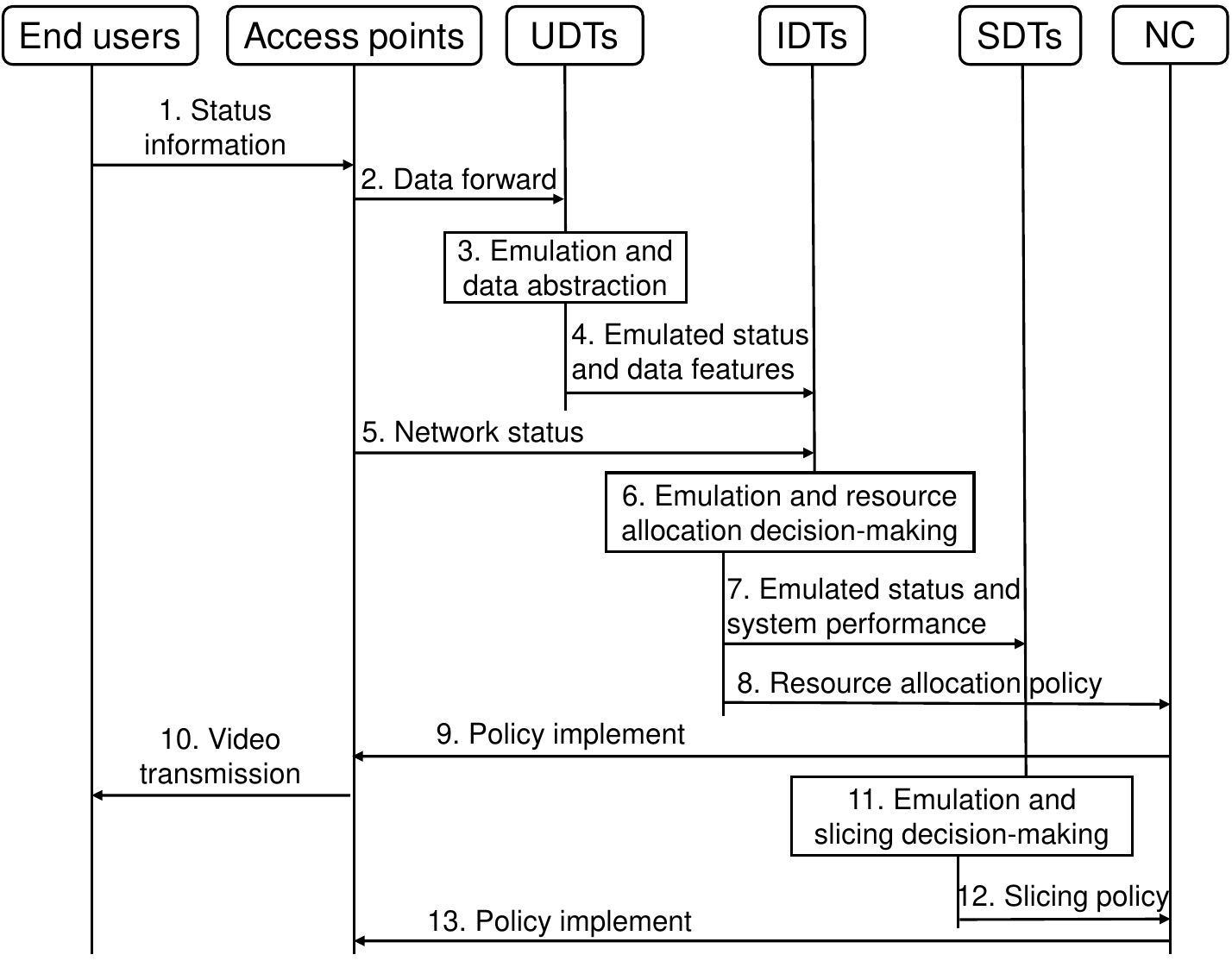}
	\caption{DT-assisted short video streaming processing procedure.}
	\label{pro}
\end{figure}

Take the short video slice as an example, the processing procedure is shown in Fig.~\ref{pro}. \ac{udt}s store users' historical status information, including channel conditions, locations, swipe timestamps, and preferences, etc. The data stored in \ac{udt}s are analyzed by embedded models to emulate user status, such as swipe behaviors, and abstract some essential user features, such as swipe probability distribution. In the small timescale, the emulated user status and abstracted user features are transferred to \ac{idt}s, integrated with network topology and performance metrics to emulate network operation status and design tailored resource allocation algorithms. The resource allocation policy will be delivered to the \ac{nc} and then implemented on the access points to facilitate real-time video transmission. In the large timescale, the emulated network operation status and system performance are further transferred to \acp{sdt} to abstract global network information for the slicing policy adjustment. The slicing policy will be delivered to the \ac{nc} and then implemented on the access points to reserve resources. 
\section{Research Challenges and Solutions}

To realize the proposed \ac{VDTNet}, some research challenges need to be addressed.

\subsection{Efficient Data Abstraction from Physical Domain}

\subsubsection{Challenge}

Intuitively, addressing the complex interplay among the physical domain, slice domain, and DT domain requires an efficient data abstraction mechanism. Since DTs need to be updated to guarantee accuracy, how to autonomously identify the types and granularities of collected data is crucial. To enhance the network management level, it is imperative to clarify what specific insights and optimizations that DTs can provide. Furthermore, the development of efficient algorithms for real-time data collection is essential, especially in ultra-low latency extreme immersive \ac{vr} scenarios.

Taking user behavior data as an example, we encounter several challenges. Initially, user behavior data, such as swipe, likes, subscribe, favorites, comments, etc., constitutes a complex user profile. It is challenging to determine the importance of different user behavior data and the data update frequency for maintaining \acp{udt}. Additionally, since user behavior data is multi-dimensional, it is challenging to discern which data dimensions can be effectively fused, and which user behavior patterns can be extracted to assist network management. Finally, user movement in \ac{vr} scenarios spans from macro gestures like head and hand motions to micro shifts in viewpoint, which requires a well-designed data abstraction algorithm to realize data synchronization with low communication overhead.

\subsubsection{Solution}
To address the complex interplay among different domains, an efficient data collection mechanism can be first developed, which relies on data importance and distribution to adjust the collected data type and granularity. Then, since network performance is usually closely related to a part of network status, a meticulously designed data fusion mechanism can be developed to abstract some distilled features from network status to facilitate network management. Finally, a lightweight semantic abstraction algorithm can be developed to capture users' real-time behaviors and network conditions, which can help reduce transmission overhead and guarantee real-time responsiveness.

Similarly, taking user behavior data abstraction as an example, we can first employ the principal component analysis technique to analyze which data types have a strong relationship with network performance, and make an adaptive data collection granularity based on data distribution variation. For data having strong statistical characteristics, \ac{dt} can generate them through the emulation module with high fidelity. Then, meticulously designed machine learning algorithms, such as autoencoders and transformers, can be leveraged to abstract and predict user behavior patterns, such as swipe probability distributions in short video streaming and \ac{roi} in \ac{vr} streaming. Finally, some lightweight feature abstraction algorithms, such as shallow \ac{vgg} and \ac{vit}, can be embedded into \ac{iot} sensors to abstract user behavior information for transmission in real time with low computing overhead.

\subsection{Comprehensive Performance Evaluation Framework} 
\subsubsection{Challenge}
The development of a comprehensive performance evaluation framework is crucial for effectively assessing DT performance in network management. Such a framework should integrate a diverse and adaptable set of \acp{kpi} to analyze DT performance. Traditional \acp{kpi} such as latency, throughput, and buffering time may not be sufficient to capture the multi-faceted nature of DT performance, which includes not only network metrics but also user behavior and \ac{qoe}. The integration of machine learning algorithms for predictive analytics, real-time data synchronization between the physical network and its DT, and edge computing for low-latency data processing adds layers of complexity to performance evaluation. Additionally, the dynamic nature of video streaming, characterized by fluctuating bit rates, stochastic video requests, and user interactivity, requires \acp{kpi} to be robust and sensitive to these fluctuations. Furthermore, the \acp{kpi} must be adaptable to different network architectures and technologies, ranging from traditional \ac{cdn} to next-generation communication infrastructures. Therefore, the challenge lies in developing a comprehensive performance evaluation framework that can holistically evaluate DT performance, taking into account the intricate interplay among the physical domain, slice domain, and DT domain.

\subsubsection{Solution}
To address this gap, a novel \ac{kpi}, termed \textit{holistic DT value}, is introduced, denoted by $V$. From the perspective of DT itself, data freshness~\cite{costa2016age} and model operation cost are very important metrics to optimize the synchronization between DTs and physical entities. From the perspective of DT impact, resource management gains, such as \ac{qos} and \ac{qoe}, directly reflect the impact of DTs on service quality and watching experience. We refer to the utility function construction method, where service delay, energy consumption, and revenue constitute the comprehensive performance metric to optimize the communication, caching, and computing resource management in vehicular networks~\cite{9491942}. Therefore, the new \ac{kpi} should integrate the key metrics to provide a holistic view of network performance, which is expressed as:
\begin{align}
V = \alpha \cdot \left( \frac{f}{\mathcal{F}} \right) + \beta \cdot Q(\myvec{C},\myvec{A},\myvec{P},\myvec{S}, \mathcal{L}) - \gamma \cdot R(\mathcal{L}),
\end{align}
where $\alpha, \beta, \gamma$ are weighting factors. Here, $f$ and $\mathcal{F}$ represent the data collection frequency and data freshness, respectively. Function, $Q$, represents the traditional \acp{kpi}, such as \ac{qos} and \ac{qoe}, which is related to communication resource scheduling matrix $\myvec{C}$, caching resource scheduling matrix $\myvec{A}$, computing resource scheduling matrix $\myvec{P}$, sensing resource scheduling matrix $\myvec{S}$, and data abstraction level $\mathcal{L}$. In the actual network optimization, we can select a part of resource scheduling decisions as the joint optimization variables to avoid dimension curse. Here, function $R$ reflects the DT model operation cost related to its data abstraction level $\mathcal{L}$, which can be quantified based on the model structure analysis.

\subsection{Adaptive DT Model Update} 
\subsubsection{Challenge}
Unlike static models, DTs must continuously evolve to reflect actual changes in both network conditions and user behaviors. This requires sophisticated machine learning algorithms capable of processing large data volumes in real time, possibly leveraging collaborative cloud-edge computing mechanism embedded with distributed learning for the low-latency DT model update. Moreover, the dynamics of video streaming, characterized by fluctuated bit rates, diverse content types, and user interactivity, also add the complexity of the DT model update. Techniques such as \ac{drl} or \ac{gan} can facilitate efficient decision-making and network emulation~\cite{8931561}, but come with their own challenges, such as the requirement of extensive training data and computing resources, and the risk of model overfitting. Furthermore, since users' network conditions, behaviors, and preferences own certain similarities, DTs can exchange part of data and learn models from each other to evolve together. This demands an efficient learning algorithm design with low computational overhead. Therefore, the challenge lies in developing adaptive DTs that can adapt to highly dynamic and complicated video streaming services.

\subsubsection{Solution}
To effectively tackle the intricate challenge of developing adaptive DT models for emerging video streaming services, a multi-layered solution that integrates distributed learning and transfer learning is put forth. In a hybrid cloud-edge computing architecture, distributed learning algorithms are employed to facilitate model split and parallel computing. The architecture allows DTs to parallelly process seamless network-related data and user-specific behavior data, which can effectively reduce service latency. Additionally, transfer learning can be particularly effective when exploiting similarities between different DTs. For instance, a model trained on one DT that has successfully been adapted to certain network conditions and user behaviors can be fine-tuned for another DT with similar conditions. This approach capitalizes on the inherent similarities between different DTs to quickly adapt to new scenarios, eliminating the requirement for extensive retraining and thereby reducing computing overhead. Moreover, a feedback loop mechanism can also be used to refine DT models, where real-time performance data are used to continuously refine the data processing algorithms and ensure that DT models remain accurate and effective in the face of evolving network dynamics and service diversity.

\section{Case Study: DT-Assisted Network Slicing for Short Video Streaming}
In this section, a case study is provided on \ac{dt}-assisted network slicing, aimed at improving the system utility consisting of user satisfaction and resource consumption.

\subsection{Considered Scenario}
 {We consider a DT-assisted multicast short video streaming (MSVS) network, which consists of two \acp{bs}, an \ac{es}, and sixty \acp{udt}. Bandwidth and computing resources are sliced (or reserved) for each multicast group to guarantee \ac{qos} requirement. Each \ac{udt} corresponds to an individual user consisting of a finite data pool and a data analysis function. Specifically, in each \ac{udt} data pool, we first simulate the user’s trajectory within the \ac{uw} campus with differentiated speed, and the user’s real-time channel condition is generated based on \texttt{propagationModel} at Matlab. Then, we employ the real-world dataset\footnote{ACM MM Grand Challenges: https://github.com/AItransCompetition/Short-Video-Streaming-Challenge/tree/main/data} to simulate the user’s swipe timestamps and preference on the sampled YouTube 8M dataset\footnote{YouTube 8M dataset: https://research.google.com/youtube8m/index.html}. The data analysis function investigates the user’s swipe timestamps to obtain a swipe probability distribution for each video type. After the construction of \acp{udt}, a \ac{drl}-based user clustering algorithm is implemented to cluster \acp{udt} into different multicast groups. \acp{udt}’ swipe timestamps and preferences are used to abstract the swipe probability distribution and recommended video list for accurate bandwidth and computing resource demand prediction in each multicast group. Based on the predicted information, the \ac{nc} can make appropriate bandwidth and computing resource reservation strategy for each multicast group to enhance the system utility.}


We propose a hybrid data-model-driven solution, where \ac{udt}s' data are analyzed by the \ac{drl}-based user clustering algorithm to update multicast groups, and the resource reservation problem is transformed into a convex problem to obtain the optimal solution \cite{huang2023digital}. For performance comparison, we adopt two schemes, i.e., 1) heuristic solution, where multicast groups are updated based on users' preferences and locations, and bandwidth and computing resource reservation is based on historical video traffic distribution; 2) optimization-based solution, where multicast groups are updated based on the density-based spatial clustering of applications with noise (DBSCAN) algorithm, and resource reservation is based on the branch- and bound-based scheduling algorithm.

\subsection{Simulation Results}

\begin{figure}[!t]
	\centering
	\includegraphics[width=0.96\linewidth]{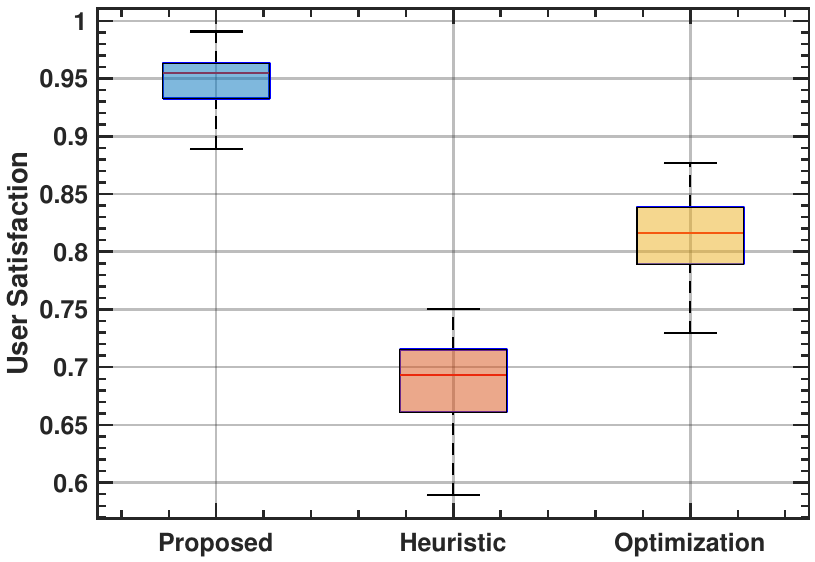}
	\caption{User satisfaction comparison.}
	\label{sat}
\end{figure}

\begin{figure}[!t]
	\centering
	\includegraphics[width=0.95\linewidth]{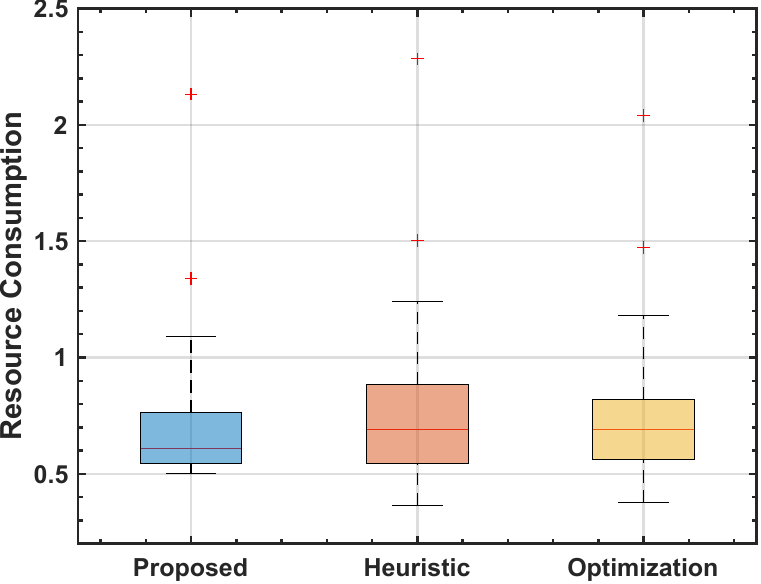}
	\caption{Resource consumption comparison.}
	\label{cons}
\end{figure}

\subsubsection{Simulation Settings}

We emulate two \ac{bs}s at the \ac{uw} campus and users' initial positions are randomly and uniformly generated around two \ac{bs}s. Each user moves along a prescribed path within the \ac{uw} campus at a speed of 2$\sim$5 $km/h$. The transmission power and noise power are set to 27 $dBm$ and -174 $dBm$, respectively. We sample 1000 short videos from the YouTube 8M dataset, which includes 8 video types, i.e., Entertainment, Games, Food, Sports, Science, Dance, Travel, and News. Each video has a duration of 15 $sec$ and is encoded into four versions by the H. 265 encoder. The detailed simulation setting can be found at~\cite{huang2023digital}.

\subsubsection{Performance Analysis}

As shown in Fig.~\ref{sat}, we present user satisfaction comparison with the help of box plot. It can be observed the proposed scheme can achieve the highest user satisfaction with the lowest fluctuation, because the \ac{dt}-based user clustering algorithm can well mine users' intrinsic correlation to accurately update multicast groups, while the convex optimization algorithm can make the optimal resource reservation strategy based on the updated multicast groups to enhance user satisfaction. The reason why user satisfaction fluctuates is due to users' time-varying resource demands and limited network resources. Finally, we present the resource consumption comparison in Fig.~\ref{cons}. It can be observed that the proposed solution can achieve
lower median, third-quartile, and maximum values compared with other schemes. Our proposed scheme
demonstrates superior performance with relatively minimal
variations, while the heuristic scheme exhibits a larger fluctuation.

\section{Open Research Issues}

\subsection{Efficient Coordination of DT Modules}

As an efficient data management platform for video streaming services, the coordination of DT modules directly influences network performance. Specifically, the development of real-time synchronization mechanisms among different DT modules is crucial for ensuring that all components are operated based on the latest data, thereby enhancing network responsiveness and DT accuracy. Furthermore, the establishment of standardized communication protocols is essential for seamless interoperability among different modules, facilitating a unified and effective system architecture. Lastly, the design of adaptive algorithms that can dynamically allocate resources among modules based on real-time performance analysis and emulation is vital for optimizing resource utilization. Therefore, it is crucial to develop an efficient coordination mechanism of DT modules to improve network performance.

\subsection{Closed-Loop Network Management}

To realize a sustainable and continuously evolving network for video streaming services, it is essential to construct an internal and external closed-loop network management system. Internally, \ac{udt}s, \ac{idt}s, and \ac{sdt}s are mainly responsible for user status analysis, network emulation, and network slicing, respectively. For instance, \ac{udt}s analyzing users' high-frequency interaction behaviors could prompt the \ac{sdt}s to reserve more resources, validated by the \ac{idt}s, which can create a self-regulating loop for optimal performance. Externally, DTs continuously monitor the physical network's status and provide useful information for network management. The physical network feeds back its actual performance data to update DT data and models. Hence, constructing a closed-loop network management system requires efficient data exchange, network emulation, data synchronization, and model update.

\subsection{Security and Privacy of DT}

While much of the current research primarily emphasizes the constructions of UDTs, IDTs, and SDTs to enhance video streaming services, there exists a notable oversight in addressing the issues of DT security and privacy protection. From the perspective of users, the urgency of data privacy escalates within the DTN4VS framework. Particularly, not only content service providers but also \acp{nc} need to gather sensitive user data, such as video preferences and locations, and the unique data collection model intensifies the complexity of effective data privacy regulation. Furthermore, the creation of DTs mandates collaboration amongst various stakeholders, which requires them to contribute their own data and analytic models~\cite{9355048}. Thus, establishing trust in such a distributed environment and protecting both data and AI model security poses significant challenges. Although current privacy-preserving techniques such as differential privacy, secure multi-party computation, and homomorphic encryption, offer potential solutions, they mandate further exploration for efficiency enhancements and tailored strategies.

\section{Conclusions}
\label{sec:Conclusion}

We have proposed the \ac{VDTNet} to realize holistic network virtualization for emerging video streaming services. Specifically, \ac{VDTNet} aims to seamlessly integrate eMBB-Plus, native AI, sensing, and network slicing through DTs to achieve efficient network management. It can further separate the resource management functions from \ac{nc}s and empower the functions with emulated data and tailored strategies, which can reduce the centralized computation burden and enhance network robustness. To supplement the \ac{VDTNet}'s functionality, we have proposed a data importance-based abstraction mechanism, a holistic DT performance evaluation metric, and a distributed transfer learning algorithm, respectively. A case study has been presented, and some open research issues have been provided for accelerating the pace of \ac{VDTNet} development.



%
\bibliographystyle{IEEEtran}
\bibliography{IEEEabrv,Ref}

\begin{thebibliography}{10}
\providecommand{\url}[1]{#1}
\csname url@samestyle\endcsname
\providecommand{\newblock}{\relax}
\providecommand{\bibinfo}[2]{#2}
\providecommand{\BIBentrySTDinterwordspacing}{\spaceskip=0pt\relax}
\providecommand{\BIBentryALTinterwordstretchfactor}{4}
\providecommand{\BIBentryALTinterwordspacing}{\spaceskip=\fontdimen2\font plus
\BIBentryALTinterwordstretchfactor\fontdimen3\font minus
  \fontdimen4\font\relax}
\providecommand{\BIBforeignlanguage}[2]{{%
\expandafter\ifx\csname l@#1\endcsname\relax
\typeout{** WARNING: IEEEtran.bst: No hyphenation pattern has been}%
\typeout{** loaded for the language `#1'. Using the pattern for}%
\typeout{** the default language instead.}%
\else
\language=\csname l@#1\endcsname
\fi
#2}}
\providecommand{\BIBdecl}{\relax}
\BIBdecl

\bibitem{report}
{Grand view research}, ``Video streaming market size, share \& trends analysis
  report by streaming type, by solution, by platform, by service, by revenue
  model, by deployment type, by user, by region, and segment forecasts,
  2023-2030,''
  \url{https://www.grandviewresearch.com/industry-analysis/video-streaming-market},
  2022, [Online; accessed 10-Oct-2023].

\bibitem{9599635}
Z.~Nadir, T.~Taleb, H.~Flinck, O.~Bouachir, and M.~Bagaa, ``Immersive services
  over {5G} and beyond mobile systems,'' \emph{IEEE Netw.}, vol.~35, no.~6, pp.
  299--306, 2021.

\bibitem{dang2020should}
S.~Dang, O.~Amin, B.~Shihada, and M.-S. Alouini, ``What should {6G} be?''
  \emph{Nat. Electron.}, vol.~3, no.~1, pp. 20--29, 2020.

\bibitem{9267778}
O.~E. Marai, T.~Taleb, and J.~Song, ``Roads infrastructure digital twin: A step
  toward smarter cities realization,'' \emph{IEEE Netw.}, vol.~35, no.~2, pp.
  136--143, Apr. 2021.

\bibitem{9915358}
Y.~Huang, Y.~Zhu, X.~Qiao, X.~Su, S.~Dustdar, and P.~Zhang, ``Toward
  holographic video communications: A promising {AI}-driven solution,''
  \emph{IEEE Commun. Mag.}, vol.~60, no.~11, pp. 82--88, 2022.

\bibitem{8258607}
C.~Timmerer, ``Immersive media delivery: Overview of ongoing standardization
  activities,'' \emph{IEEE Commun. Stand. Mag.}, vol.~1, no.~4, pp. 71--74,
  2017.

\bibitem{9749222}
W.~Wu, C.~Zhou, M.~Li, H.~Wu, H.~Zhou, N.~Zhang, X.~Shen, and W.~Zhuang,
  ``{AI}-native network slicing for {6G} networks,'' \emph{IEEE Wirel.
  Commun.}, vol.~29, no.~1, pp. 96--103, 2022.

\bibitem{9941042}
C.~Chen, H.~Song, Q.~Li, F.~Meneghello, F.~Restuccia, and C.~Cordeiro,
  ``{Wi-Fi} sensing based on {IEEE} 802.11bf,'' \emph{IEEE Commun. Mag.},
  vol.~61, no.~1, pp. 121--127, 2023.

\bibitem{9363029}
C.~B. Barneto, S.~D. Liyanaarachchi, M.~Heino, T.~Riihonen, and M.~Valkama,
  ``Full duplex radio/radar technology: The enabler for advanced joint
  communication and sensing,'' \emph{IEEE Wirel. Commun.}, vol.~28, no.~1, pp.
  82--88, 2021.

\bibitem{9651548}
X.~Shen, J.~Gao, W.~Wu, M.~Li, C.~Zhou, and W.~Zhuang, ``Holistic network
  virtualization and pervasive network intelligence for {6G},'' \emph{IEEE
  Commun. Surv. Tut.}, vol.~24, no.~1, pp. 1--30, 2022.

\bibitem{costa2016age}
M.~Costa, M.~Codreanu, and A.~Ephremides, ``On the age of information in status
  update systems with packet management,'' \emph{IEEE Trans. Inf. Theory},
  vol.~62, no.~4, pp. 1897--1910, 2016.

\bibitem{9491942}
L.~Xu, Z.~Yang, H.~Wu, Y.~Zhang, Y.~Wang, L.~Wang, and Z.~Han, ``Socially
  driven joint optimization of communication, caching, and computing resources
  in vehicular networks,'' \emph{IEEE Trans. Wirel. Commun.}, vol.~21, no.~1,
  pp. 461--476, 2022.

\bibitem{8931561}
Y.~Hua, R.~Li, Z.~Zhao, X.~Chen, and H.~Zhang, ``{GAN}-powered deep
  distributional reinforcement learning for resource management in network
  slicing,'' \emph{IEEE J. Sel. Areas Commun.}, vol.~38, no.~2, pp. 334--349,
  2020.

\bibitem{huang2023digital}
X.~Huang, W.~Wu, S.~Hu, M.~Li, C.~Zhou, and X.~Shen, ``Digital twin based
  user-centric resource management for multicast short video streaming,''
  \emph{arXiv preprint arXiv:2308.08995}, 2023.

\bibitem{9355048}
X.~Shen, C.~Huang, D.~Liu, L.~Xue, W.~Zhuang, R.~Sun, and B.~Ying, ``Data
  management for future wireless networks: Architecture, privacy preservation,
  and regulation,'' \emph{IEEE Netw.}, vol.~35, no.~1, pp. 8--15, 2021.

\end{thebibliography}
%
%
%
\begin{IEEEbiographynophoto}{Xinyu Huang} (S'21) received the B.E. and M.S. degrees from Xidian University and Xi'an Jiaotong University, Xi’an, China, in 2018 and 2021, respectively. He is working toward the Ph.D. degree in Electrical and Computer Engineering at the University of Waterloo, Waterloo, ON, Canada. His research interests include digital twins, generative AI, and network resource management.
\end{IEEEbiographynophoto}
\begin{IEEEbiographynophoto}{Haojun Yang} (M'21) received the B.S. degree in communication engineering and the Ph.D. degree in information and communication engineering from Beijing University of Posts and Telecommunications (BUPT), Beijing, China, in 2014 and 2020, respectively. He is currently a Postdoctoral Fellow with the Department of Electrical and Computer Engineering, University of Waterloo, Waterloo, Canada. His research interests include ultra-reliable and low-latency communications, resource management and vehicular networks.
\end{IEEEbiographynophoto}
\begin{IEEEbiographynophoto}{Shisheng Hu} (S'19) received the B.Eng. degree and 	the M.A.Sc. degree from the University of Electronic Science and Technology of China (UESTC), Chengdu, China, in 2018 and 2021, respectively. He is currently working toward the Ph.D. degree	with the Department of Electrical and Computer Engineering, University of Waterloo, Waterloo, ON, Canada. His research interests include AI for wireless networks and networking for AI.
\end{IEEEbiographynophoto}
\begin{IEEEbiographynophoto}{Xuemin (Sherman) Shen} (M'97–SM'02–F'09) received the Ph.D. degree in electrical engineering from Rutgers University, New Brunswick, NJ, USA, in 1990. He is a University Professor with the	Department of Electrical and Computer Engineering, University of Waterloo, Canada. His research focuses on network resource management, wireless network security, Internet of Things, AI for networks, and vehicular networks. Dr. Shen is a registered Professional Engineer of Ontario, Canada, an Engineering Institute of Canada Fellow, a Canadian Academy of Engineering Fellow, a Royal Society of Canada Fellow, a Chinese	Academy of Engineering Foreign Member, and a Distinguished Lecturer of	the IEEE Vehicular Technology Society and Communications Society.
\end{IEEEbiographynophoto}

\end{document}